# IDENTIFICATION OF PROMOTED ECLIPSE UNSTABLE INTERFACES USING CLONE DETECTION TECHNIQUE

Simon Kawuma[1] and Evarist Nabaasa[2]

[1]Department of Computer Engineering, Mbarara University of Science and Technology, Mbarara, Uganda

[2]Department of Computer Science, Mbarara University of Science and Technology, Mbarara, Uganda

*ABSTRACT*

*The Eclipse framework is a popular and widely used framework that has been evolving for over a decade. The framework provides both stable interfaces (APIs) and unstable interfaces (non-APIs). Despite being discouraged by Eclipse, client developers often use non-APIs which may cause their systems to fail when ported to new framework releases. To overcome this problem, Eclipse interface producers may promote unstable interfaces to APIs. However, client developers have no assistance to aid them to identify the promoted unstable interfaces in the Eclipse framework. We aim to help API users identify promoted unstable interfaces. We used the clone detection technique to identify promoted unstable interfaces as the framework evolves. Our empirical investigation on 16 Eclipse major releases presents the following observations. First, we have discovered that there exists over 60% non-API methods of the total interfaces in each of the analyzed 16 Eclipse releases. Second, we have discovered that the percentage of promoted non-APIs identified through clone detection ranges from 0.20% to 10.38%.*

## 1. INTRODUCTION

Today, developers build their software application on top of frameworks. This has advantages such as the production of good quality software and reduced software development time since the application developer reuses the functionality provided by the framework [1] and increased productivity [2]. Eclipse is a widely used and adopted framework. It is a large and complex open source software system used by thousands of software developers. Eclipse framework provides two types of Application Programming interfaces to application developers namely; stable interfaces (APIs) and unstable Interfaces (non-APIs). Eclipse framework developers discourage the use of non-APIs because they are immature, unsupported, undocumented and subject to be changed or removed from the framework without notice [3].

Despite being discouraged, the usage of non-APIs is common. From previous studies, Businge et al. made a number of observations: First, they found out that about 44% of 512 analyzed Eclipse plug-in use non-APIs [4]. Second, they discovered that as Eclipse interface providers state, APIs are indeed stable as they do not cause compatibility failures to applications that solely depend on them [5]. Third, they also observed that indeed non-APIs cause compatibility failures to applications that use them in new framework releases[5] [6]. Fourth, they discovered that the major reason why application developers use non-APIs is that they claim that they cannot find APIs with the functionality they require among stable APIs and therefore compelled to use non-APIs [7]. Indeed, Kawuma et al. showed that less than 1% APIs offer the same or similar functionality as non-APIs in the 18 Eclipse major releases[8].

Furthermore, in our previous study, Kawuma et al. observed that there were twice as many fully qualified non-API methods compared to APIs [8]. Still, from the same, we also observed that both non-API and API methods are growing although the rate at which new non-API methods are introduced in new Eclipse version is much higher compared to the rate at which APIs are introduced. This implies that more new non-API files are introduced instead of graduating the already existing old non-API file to API files. It is possible that there is a very low promotion rate of non-APIs to APIs. To overcome the problem of compatibility failures caused by using unstable interfaces, interface producers should expedite graduation of unstable interfaces to stable interfaces, however; application developers who use the framework have no assistance to identify the promoted unstable interfaces. To this end, we propose to use the clone detection technique to identify and recommended the promoted unstable interfaces to Application developers. We carried out an investigation on 16 Major Eclipse releases; first to establish the number of non-APIs in different Eclipse releases. Secondly to establish the number of unstable interfaces promoted to stable interfaces. We aim to recommend the identified promoted unstable interfaces to application developers. We formulate two research questions to guide the study as follows:

**RQ1: What percentage of non-API methods exists in the different Eclipse releases?** From the analysis, we have discovered that there exist over 60% unstable interfaces in all the analyzed 16 Eclipse releases.

**RQ2: What is the percentage of non-API methods promoted to API methods?** We have discovered that the percentage of identified promoted unstable interfaces for all the analyzed 16 Eclipse releases ranges from 0.20% to 10.38 %. The identify promoted unstable interface would be recommended to the application developer.

The remainder of this paper is organized as follows: Section 2 presents the background information on Eclipse interfaces and software clones. Section 3 discusses the experimental setup of our study. Section 4 presents the results and findings of our study. Section 5 presents threats to the validity of our study. Finally, Section 6 presents the conclusion and future work**.**

## 2. BACKGROUND

### 2.1. ECLIPSE UNSTABLE INTERFACES (NON-APIS)

Non-APIs are internal implementation artifacts that according to Eclipse naming convention [9] are found in packages with the substring internal in the fully qualified name. These internal implementation artifacts include public Java classes or interfaces, or public or protected methods, or fields in such a class or interface. Usage of non-APIs is strongly discouraged since they may be unstable. Eclipse clearly states that clients who think they must use these non-APIs do it at their own risk as non-APIs are subject to arbitrary change or removal without notice. Eclipse does not usually provide documentation and support to these non-APIs.

### 2.2. ECLIPSE STABLE INTERFACES (APIS)

These are public Java classes or interfaces that can be found in packages that do not contain the segment internal in the fully qualified package name, a public or protected method, or field in such a class or interface [9]. Eclipse states that the APIs are considered to be stable and therefore can be used by any application developer without any risk. Furthermore, Eclipse also provides documentation and support for these APIs. APIs which are well designed ease program comprehension, reduced software complexity and maintainability [10].

### 2.3. UNSTABLE INTERFACES USAGE AND PROMOTION

Businge et al. [4-7]performed several empirical investigations on the usage of unstable, interfaces of the Eclipse platform, they discovered that 44% of 512 analyzed eclipse plugins depended on internal interfaces. Hora et al. [11] replicated their analysis at an ultra-large scale level and found out that 23.5% of 9702 Eclipse client project on Github depended on internal interfaces. Wu.W et al. [12] analyzed and classified API changes and usage in 22 framework releases from the Apache, ecosystems and their client programs. They discovered that framework APIs are used on average in 35% of client classes and interfaces and that about 11% of APIs usages could cause ripple effects in client programs when these APIs change. Mastrangelo et al. [13] discovered that client projects often use internal interfaces provided by JDK. Hani Abdeen et al. [10] evaluated the impact of interface clone on interface design quality and discovered that interface clones are reliable symptoms of poor interface design.

Andre Hora et al. [11] investigated the transition from internal to public interfaces. They carried out their investigation on Eclipse (JDT), JUnit, Hibernate, jB-PM, and ElasticSearch. They discovered that 7% of 2,277 of internal interfaces are promoted to public interfaces. They also found that the promoted interfaces have more clients. Furthermore, they predicted internal interface promotion with a precision and recall between 50%-80% and 26%-82% respectively. Using their predictor on the analyzed systems, they found 382 public interface candidates. Hora et al. [14, 15] propose tools to keep track of API evolution by mining fine-grained code changes Comparing several studies with ours, we use clone detection technique to identify the promoted unstable interfaces.

### 2.4. CLONE TERMINOLOGY

Software clone detection is a well-established research area [16] in the remainder of the section we briefly introduce notions related to clone detection.

**Code Fragment:** A code fragment is a sequence of code lines with or without comments[16]. A code fragment is identified by its file name and begin-end line numbers in the original code base.

**Code Clone:** A code fragment CF2 is a clone of another code fragment CF1 if they are similar by some given definition of similarity. (CF1; CF2) then form a clone pair. If multiple fragments are similar, they form a clone class or clone group[16]

**Clone Types:** Depending on the definition of similarity, different clone types can be distinguished. In this study we consider code clones of Types I, II and III as defined by Roy and Cordy as follows:

- Type-I clones are identical code fragments except for variations in white space, layout, and comments.

- Type-II clones are structurally and syntactically identical except for variations in identifiers, literals, types, layout, and comments.

- Type III code clones are copies with further modifications, statements can be changed, added, or removed in addition to variations in identifiers, literals, types, layout, and comments.

## 3. EXPERIMENTAL SETUP

This section details the experimental setup of how the data for the research questions was collected.

### 3.1. DATA COLLECTION

In this section, we explain the data sources of our study. Our study is based on all the 16 Eclipse SDK major releases from Eclipse project Archive website [17] until Eclipse 4.6. Table 1 presents all the Eclipse major releases with their corresponding release dates.

Table 1: Eclipse major releases and their corresponding release dates.

| Eclipse Releases | Release Date | Eclipse Releases | Release Date |
|---|---|---|---|
| E-1.0 | 07-Nov-01 | E-3.5 | 11-Jun-09 |
| E-2.0 | 27-Jun-02 | E-3.6 | 08-Jun-10 |
| E-2.1 | 27-Mar-03 | E-3.7 | 13-Jun-11 |
| E-3.0 | 25-Jun-04 | E-4.2 | 08-Jun-12 |
| E-3.1 | 27-Jun-05 | E-4.3 | 05-Jun-13 |
| E-3.2 | 29-Jun-06 | E-4.4 | 06-Jun-14 |
| E-3.3 | 25-Jun-07 | E-4.5 | 03-Jun-15 |
| E-3.4 | 17-Jun-08 | E-4.6 | 06-Jun-16 |

We used the NiCad clone detection tool [18] to extract the data for all RQs. In addition to the NiCad being suitable for extracting the data for our research questions, the tool has also been extensively validated by previous researchers [19, 20].

## 3.2 DATA EXTRACTION FOR A NUMBER OF NON-APIS IN DIFFERENT ECLIPSE RELEASES

When an Eclipse release source directory is subjected to the NiCad tool to extract clones, the tool first extracts and generates an XML output report file containing a list of all methods in any given Eclipse release. For example in Figure 1 below, we show a representation of the XML report generated by subjecting Eclipse-1.0 to the tool.

```
<source>
    public void md1 (Cl c1) {File="eclipse-1.0/org/eclipse/p1/sp1/pk1/F1.java"}
</source>
<source>
    void m2 (){File="eclipse-1.0/org/eclipse/p3/internal/sp3/pk3/F3.java"}
</source>
<source>
    public C1 m3 (){File="eclipse-1.0/org/eclipse/p5/internal/sp5/pk5/F5.java"}
</source>
<source>
    public C1 m4 (){File="eclipse-1.0/org/eclipse/p6/sp6/pk6/F6.java"}
</source>
```

Figure 1: Nicad Extracted Method Report

We used this report to obtain the total number of non-API methods in each Eclipse release. To determine the percentage of non-APIs in a given eclipse release, we counted the number of methods with a sub-string internal in the XML report divided by the total number of methods in that particular Eclipse release.

## 3.4. DATA EXTRACTION FOR PROMOTED NON-API

To extract data for RQ2 we used NiCad cross-clone detection tool, a component of the NiCad clone detection tool [18], to extract Type-I, Type-II and Type-III clones that exist between two Eclipse releases E_old and E_new. The tool takes as input the two source directories of the Eclipse releases and produces an output report that contains only clones of fragments of E_old in E_new. We run the source files using the following NiCad configuration files type1.cfg, type2.cfg and type3.cfg for Type-I, Type-II and Type-III clones respectfully. When carrying out the experiments, we are aware that in the new Eclipse releases, there exist non-APIs which were introduced in the earlier Eclipse releases. Therefore, before carrying out the NiCad cross-clone, we first eliminate the old non-APIs in the new Eclipse release by locating the unchanged non-API in E_new that were previously introduced E_old. This leaves only newly introduced non-APIs in E_new.

For example, if we are going to carry out a NiCad cross-clone between Eclipse-3.0 and the later Eclipse-3.1, 3.2, 3.3, etc, this is how we would proceed with the experiments: 1) We extract sets of all the fully qualified names of the non-API methods in earlier Eclipse releases (i.e., 1.0, 2.0, and 2.1). 2) We then extract a set with all the fully qualified names of the non-APIs methods in the current Eclipse release of interest Eclipse 3.0.  3) Using set differencing, we eliminate all the old non-APIs methods of all earlier releases from Eclipse 3.0.  4) After obtaining the newly introduced non-API methods, we renamed this set with a new name called Eclipse-3.0-new.  We then carry out NiCad cross-clone between Eclipse-3.0-new and the later Eclipse releases and we examine the existence of clones in each of the successive Eclipse releases.   In figure 2 below, we show a representation of the XML clone report generated by subjecting Eclipse-1.0 and Eclipse-4.6 to the NiCad cross-clone tool.

```
<clones>
  <clone nlines="74" similarity="100">
    public void md1(Cl c1) {<source File="eclipse-1.0/org/eclipse/p1/sp1/pk1/F1.java">
    public void md1(Cl c1) {<source File="eclipse-4.6/org/eclipse/p2/sp2/pk2/F2.java">
  </clone>
  <clone nlines="79" similarity="100">
    void m2(){<source File="eclipse-1.0/org/eclipse/p3/internal/sp3/pk3/F3.java">
    void m2(){<source File="eclipse-4.6/org/eclipse/p4/internal/sp4/pk4/F4.java">
  </clone>
  <clone nlines="90" similarity="100">
    public C1 m3(){<source File="eclipse-1.0/org/eclipse/p5/internal/sp5/pk5/F5.java">
    public C1 m3(){<source File="eclipse-4.6/org/eclipse/p6/sp6/pk6/F6.java">
  </clone>
</clones>
```

Figure 2:  NiCad Cross Clone Report

From figure 2, one can tell that a method in a clone pair is a non-API if it has a substring internal in the source file path. For example, the first and second clone pairs in figure 1 belong to API and non-APIs respectively. Whereas the third clone pair have the first and its second source file path originating from non-API and API respectively. It should be noted that in the last clone pair, a non-API from Eclipse-1.0 was promoted to API in Eclipse-4.6. To determine the percentage of non-APIs that were promoted to APIs, we count the number of clone pairs in the output report with the similar characteristic as the last clone pair in figure 1 divided by the total number of non-API methods in the old.

## 4. RESULTS AND DISCUSSION

In this section, we present the results and analysis of the extracted data in Section 2 to address RQ1 and RQ2.

## 4.1. PERCENTAGE OF NON-APIS IN DIFFERENT ECLIPSE RELEASES

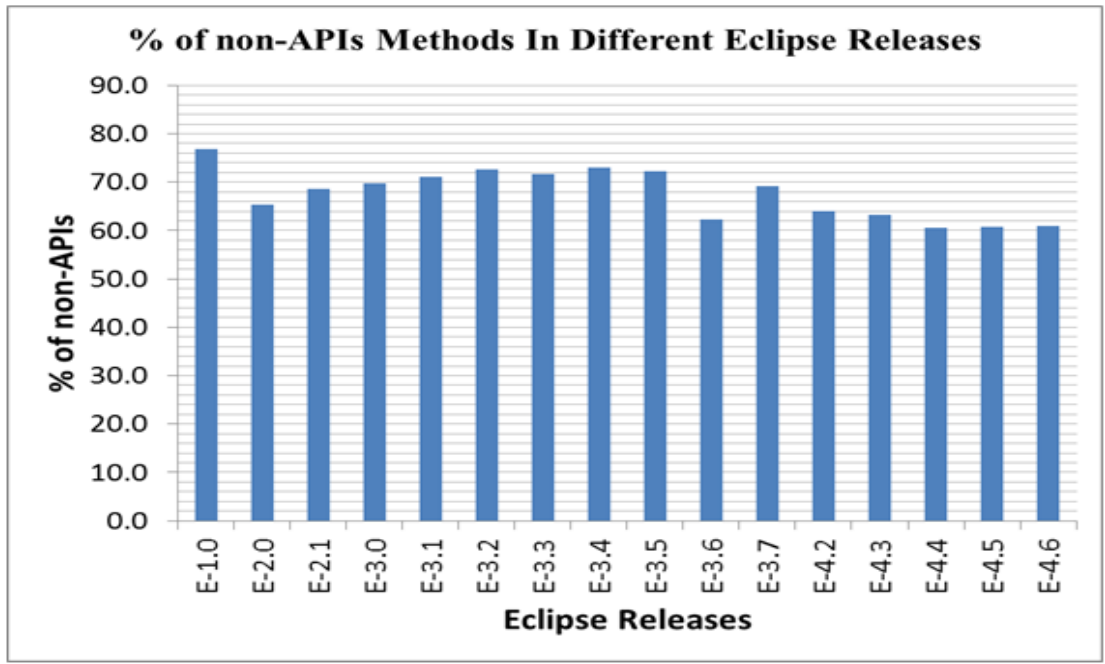

Figure 3: Percentage of non-APIs in the different Eclipse releases

Figure 3 shows the percentage of non-APIs present in each Eclipse release. We observe from Figure 3 that we have over 60% non-API methods of the total interfaces in each of the analyzed 16 Eclipse releases. The Eclipse framework has evolved for over 15 years producing a major release every year with the aim of providing new interfaces and improving on the quality of existing interfaces to application developers, one would expect that interfaces labeled unstable to be at least reducing. However, from our observations, the rate at which the non-API methods are reducing seems to be very slow. The observation tells us that this is a considerably a large amount of non-API methods in the different Eclipse releases.

From the discussion of results in this section, we have indeed confirmed our earlier preliminary observation in Kawuma et al [8], that indeed the introduction of non-APIs in the Eclipse releases is much higher than that of APIs. The findings of this research question gave us a very good motivation to investigate research question RQ2 as presented in the next section.

## 4.2. PERCENTAGE OF NON-APIS PROMOTED TO APIS

| Eclipse Versions | E 1.0 | E 2.0 | E 2.1 | E 3.0 | E 3.1 | E 3.2 | E 3.3 | E 3.4 | E 3.5 | E 3.6 |
|---|---|---|---|---|---|---|---|---|---|---|
| New | 30766 | 26500 | 12549 | 32634 | 18037 | 21245 | 12772 | 23794 | 14398 | 9482 |
| E 2.0 | 2.10 | | | | | | | | | |
| E 2.1 | 2.91 | 2.71 | | | | | | | | |
| E 3.0 | 0.68 | 0.22 | 0.20 | | | | | | | |
| E 3.1 | 3.50 | 3.59 | 3.75 | 4.42 | | | | | | |
| E 3.2 | 3.39 | 3.92 | 4.03 | 4.91 | 4.19 | | | | | |
| E 3.3 | 3.36 | 4.10 | 4.27 | 4.94 | 4.85 | 7.96 | | | | |
| E 3.4 | 3.70 | 4.57 | 4.43 | 5.51 | 5.44 | 8.57 | 6.25 | | | |
| E 3.5 | 4.12 | 5.03 | 4.81 | 5.97 | 5.91 | 9.04 | 6.68 | 3.18 | | |
| E 3.6 | 4.96 | 6.06 | 5.85 | 6.78 | 6.65 | 10.38 | 7.36 | 3.57 | 5.93 | |
| E 3.7 | 0.69 | 0.91 | 1.03 | 0.95 | 0.78 | 1.18 | 1.06 | 0.45 | 0.62 | 0.83 |
| E 4.2 | 0.99 | 1.54 | 1.12 | 1.77 | 1.92 | 2.80 | 2.32 | 1.17 | 1.87 | 1.97 |
| E 4.3 | 1.00 | 1.55 | 1.14 | 1.74 | 1.97 | 2.84 | 2.66 | 1.30 | 2.18 | 1.91 |
| E 4.4 | 1.07 | 1.62 | 1.25 | 1.81 | 1.99 | 2.97 | 2.90 | 1.31 | 2.39 | 1.96 |
| E 4.5 | 1.03 | 1.55 | 1.38 | 1.85 | 1.70 | 2.78 | 2.76 | 1.30 | 2.30 | 2.05 |
| E 4.6 | 0.97 | 1.51 | 1.33 | 1.80 | 1.62 | 2.48 | 2.64 | 1.27 | 2.21 | 2.04 |
| Min | 0.68 | 0.22 | 0.20 | 0.95 | 0.78 | 1.18 | 1.06 | 0.45 | 0.62 | 0.83 |
| Max | 4.96 | 6.06 | 5.85 | 6.78 | 6.65 | 10.38 | 7.36 | 3.57 | 5.93 | 2.05 |
| Average | 2.30 | 2.78 | 2.66 | 3.54 | 3.37 | 5.10 | 3.85 | 1.69 | 2.50 | 1.79 |

Table 2 present the percentage of promoted non-APIs in successive Eclipse releases with respect to the newly introduced non-APIs.

The Promoted non-APIs in Table 2 were identified through cross clone detection. The first row of Table 2 shows the different Eclipse old releases E_old where we carried NiCad cross clone detection with the corresponding successive Eclipse new releases E_new in the first column. The second row labeled New shows the number of newly introduced non-API methods in the different Eclipse releases e.g. E-3.0 has a total of 32,634 newly introduced non-API methods at the time of its release.

The numbers presented in the matrix of Table 2 show the percentage of promoted non-APIs of old Eclipse release (i.e. E-1.0 to E-3.6) in successive new Eclipse releases in the first column. Each value in the matrix of Table 2 show the sum of Type1, Type2 and Type3 promoted non-APIs. For example, the values in column E-3.0 show the percentages of non-API methods of 32,634 that were promoted to API methods in successive Eclipse releases (E-3.1, E-3.2, E-3.3, etc.). Looking at the value in cell (column E-3.2, row E-3.6) = 10.38 means that 10.38% of 21,245 non-API methods introduced in E-3.2 were promoted to API method in E-3.6. The last three rows of Table 2 present the descriptive statistics of promoted non-API to API methods. Looking at the last three rows specifically rows labeled min and max, the percentage of promoted non-API methods identified through cross clone detection range from 0.20% to 10.38% for all studied Eclipse releases.

The low percentage values in Table 2 indicate that there are few non-API methods that are promoted to API. We found out in section 4.1 that the rate at which the non-API methods are reducing was very slow across different Eclipse releases so one would expect to see high promotion rate of the already existing non-API methods however the rate at which non-API methods are promoted to API method is slow. Focusing on the value in the cell (column-E-1.0, row-4.6)=0.97% in Table 2 and looking at the release date of Eclipse-1.0 (Nov 2001) and Eclipse-4.6 (June 2016) in Table 1 the value in the later cell indicates that 0.97% of 30,766 of the newly introduced non-APIs was promoted after 15 years during the evolution of Eclipse Framework. A similar explanation can be used to describe how long it takes to promote non-APIs of a given Eclipse releases.

## 5. THREATS TO VALIDITY

### 5.1. CONSTRUCT VALIDITY

The methodology used to determine the number of non-API methods and promoted non-APIs in Eclipse releases may be slightly subjected to construct validity. The reason being that we only used methods in our computations yet there are other objects we have ignored, for example, variable declarations. However, it should be noted that users of the framework call methods while developing their applications

### 5.2. INTERNAL VALIDITY

Internal validity related to the NiCad tool used to extract the data used in our experiments. It is possible that results could differ if a different tool was used. Like any other static analysis tool, the NiCad tool we used does not have a 100% precision. However other studies that have compared NiCad tool with other clone detection tools have observed that NiCad tool has the highest precision and recall of any existing code clone detector tools [19, 20].

### 5.3. EXTERNAL VALIDITY

We focused on the analysis of widely adopted and large-scale Eclipse framework. It is therefore, a credible and representative case study. The framework is open source and thus its source code is easily accessible. Our findings can be directly generalized to other systems, for example, the Netbeans framework

## 6. CONCLUSIONS

In this research study, we have investigated Eclipse unstable (non-API) interfaces for the entire 16 major releases Using NiCad cross-clone detection tool. First, we have discovered that there exists over 60% non-API methods of the total interfaces in each of the analyzed 16 Eclipse releases. We have also discovered that in all the analyzed Eclipse releases, the percentage of promoted non-API identified through cross clone detection ranges from 0.20% to 10.38% for all studied Eclipse Release. The low percentage values of these results indicate that there is a slow rate at which non-APIs are promoted to APIs methods between Eclipse releases.

In our follow up study, we would like to carry out a qualitative survey with the Eclipse framework developers to establish why some non-API methods were promoted and why it takes longer to promote some non-APIs. Furthermore, we will also build an Eclipse plugin that will be integrated into Eclipse IDE to aid application developers to identify Promoted non-API interfaces.

**ACKNOWLEDGMENTS**

We would like to thanks Dr. James R. Cordy for providing the latest version of the NiCad tool and also answering our inquiries regarding NiCad.

**AUTHORS**

Simon Kawuma received his MSc degree in Embedded Computing System at Norwegian University of Science and Technology Norway and University of Southampton United Kingdom in 2013. He is currently a Ph.D. student at Mbarara University of Science and Technology. His research interests include software engineering, software maintenance, and Evolution.

Evarist Nabaasa is a senior lecturer and Dean Faculty of Computing and Informatics at Mbarara University of Science and Technology. He earned his Ph.D. in Computer Science from Mbarara University of Science and Technology in 2015. His research interests include software engineering, Computer Networks, and algorithms.